\documentclass[epj]{svjour}
\usepackage{amsmath}
\usepackage{amssymb}
\usepackage{graphicx}

\begin{document}

\title{Bounds and optimisation of orbital angular momentum bandwidths within parametric down-conversion systems}
\author{F.~M.~Miatto\inst{1}, D.~Giovannini\inst{2}, J.~Romero\inst{2}, S.~Franke-Arnold\inst{2}, S.~M.~Barnett\inst{1}, M.~J.~Padgett\inst{2}}
\institute{SUPA and Department of Physics, University of Strathclyde, Glasgow G4 0NG, Scotland, U.K.\and School of Physics and Astronomy, SUPA, University of Glasgow, Glasgow G12 8QQ, U.K.}
\abstract{
The measurement of high-dimensional entangled states of orbital angular momentum prepared by spontaneous parametric down-conversion can be considered in two separate stages: a generation stage and a detection stage. Given a certain number of generated modes, the number of measured modes is determined by the measurement apparatus. We derive a simple relationship between the generation and detection parameters and the number of measured entangled modes.}
\authorrunning{F. Miatto et al.}
\titlerunning{Bounds of OAM bandwidths within PDC Systems}
\date{Received: date / Revised version: \today}
\maketitle

\section{Introduction}
Entangled states are a distinctive feature of quantum mechanics. Their use can lead to important technological advances in communication, security and, ultimately, computing \cite{Neilsen}. Entanglement in a high-dimensional Hilbert space means a high effective number of entangled modes that can be used to achieve a high shared information \cite{Molina-Terriza:2008}. It is therefore of great importance to choose the proper basis in which to detect entangled modes. The Schmidt basis is  that which yields the maximum shared information \cite{BarnettQI}. For most entangled states, however, the elements of the Schmidt basis cannot readily be measured, perhaps because the size of the components in the detection apparatus do not match any possible detection mode \cite{Miatto:2011b}.

It is well-known that light can carry orbital angular momentum (OAM) and that this property is associated with a helical phase front \cite{AllenOAM} (and papers reprinted therein).  Optical modes carrying OAM include the Laguerre-Gaussian modes \cite{Siegman} and also the Bessel beams.
Of central interest to us in this paper is the fact that photon pairs produced in spontaneous parametric down-conversion (SPDC) are naturally entangled in their OAM \cite{MairNat01,Sonja:2002,Torres:2003}.  One clear manifestation of this entanglement is the existence of an EPR paradox for
the OAM and its conjugate quantum variable, the azimuthal angle \cite{Leach:2010,BarnettPRA90,Pors-Miatto:2011}.  

The OAM is conserved in the down-conversion process and hence for a Gaussian ($\ell = 0$) pump, the
OAM of the signal and idler fields are perfectly anticorrelated.  There are also correlations
on the radial direction (as quantified, for the Laguerre-Gaussian modes, by a radial index $p$) \cite{Miatto:2011} but these will not concern us in this paper.  Our central concern will be the number of entangled lowest order ($p=0$) Laguerre-Gaussian modes generated in a down-conversion experiment. The typical setup that we consider is a type-I or type-II, degenerate SPDC setup. We work in the regime of undepleted pump and we neglect eventual anisotropies of the down-converted beams.

We find that, for any given set of generation parameters (pump waist $w_p$, wavelength $\lambda$, crystal length $L$) the detection apparatus can be prepared in a way that maximises the measured number of entangled modes and that two important parameters are $\gamma$, the ratio of the width of the pump beam to the width of the detection modes, and $L_R$, the length of the crystal normalised to the Rayleigh range of the pump beam:

\begin{equation}
\gamma_{s,i} = \frac{w_p}{w_{s,i}} \quad \mathrm{and} \quad L_R=\frac{L}{z_R}\, .
\label{LRdef}
\end{equation}
Where the Rayleigh range is $z_R=\frac{\pi w_p^2}{\lambda}$.
In this paper we assume that the signal and idler modes have the same width so that $w_s = w_i$ and $\gamma_s = \gamma_i = \gamma$.

The precise calculation of $w_{s,i}$ depends upon the details of the detection system. Our analysis can be applied if the back-projected detection mode size, $w_{i,s}$, is approximately $\ell$-independent over the range of OAM of interest, and if the modes with $p\neq0$ couple only weakly with the fundamental mode of the fibre that carries the signal to the coincidence counter.

We investigate the $L_R$ dependence of the OAM bandwidth, while recognising that many experiments operate in a regime where $L_R\ll1$ \cite{MairNat01,Jack:2010,Oemrawsingh2005,DadaNP11,kwiat2005}.
In the short crystal limit and near to collinearity the familiar sinc phase can be dropped \cite{Saleh}.
One can then obtain an analytical form for the down-converted state \cite{Miatto:2011,BenninkPRA10} and its extension to non-Gaussian pump beams \cite{YaoNJP11}.  Our aim in this paper is to go beyond these existing analyses and to explore regimes in which the sinc phase matching term becomes significant, which leads to the exact analytical expression \eqref{analytic} and to the characterisation of the detection parameters. We present both an analytical treatment and also a simple geometrical argument for our results.

The second section of the paper specifies the definitions of the various bandwidths which are used. The third section contains the analytical approach to calculate the projection amplitudes. The fourth section contains the geometrical approach to calculate a simple formula that gives the measurement bandwidth. The fifth section contains the interpretation of the results and the conclusions.

\section{Definition of bandwidths}
For a distribution of probabilities, in our case for the OAM of the signal or idler photon in SPDC,
we can define a number of statistical measures.  For high-dimensional entanglement we require as many modes as possible
to contribute to the state and, moreover, for these to contribute strongly, that is to have a significant probability.
A simple and convenient measure of this quantity is the Schmidt number \cite{LawPRL04,PorsPRL:2008}:
\begin{align}
 K(\{p_i\}):=\frac{1}{\sum_ip_i^2}\, ,
\end{align}
where the probabilities $\{p_i\}$ are, in our case, those for each of the OAM modes.
The measure $K$ gives the effective number of contributing modes and hence the effective dimensionality of the system.  
In experiments, it is typical to quote the full-width at half maximum as the measure of the bandwidth (FWHM) so as to include only modes that are well above the noise floor. FWHM should not be confused with $K$. For simple, symmetrical and single-peaked probability distributions, the Schmidt number provides a convenient measure of the bandwidth.  The precise relationship between the Schmidt number and the FWHM depends upon the detailed shape of the distribution but typical of our systems is that the $K$ exceeds the FWHM, see figure \ref{example}.  For a distribution like this we can define an effective 
range of modes contributing to the state ranging from $\ell_\mathrm{max}$ to $\ell_\mathrm{min}=-\ell_\mathrm{max}$ such that $K=1+2|\ell_\mathrm{max}|$. 

The \emph{generation bandwidth} is the effective number of entangled modes generated in the SPDC process. As it does not depend on the detection apparatus, it is a function only of the crystal length and of the size of the pump beam, combined into the quantity $L_R$, defined in eq.~\eqref{LRdef}. This bandwidth can be thought of as the dimensionality of the entanglement in OAM and can be calculated through the Schmidt decomposition of the SPDC state \cite{Miatto:2011b}. More on the generation bandwidth is detailed in its derivation, in section 4.1.

The \emph{measurement bandwidth} represents the number of modes that a detector will measure in an experiment and depends on both the generated modes and on the overlap of these with the detection modes. In doing so, we need to consider the optics used to image the light onto the detectors and any restriction arising from this, such as a restriction to $p=0$ 
Laguerre-Gaussian modes. The overlap between the generated modes and the back-projected detection modes needs to be maintained both in the image plane and in the far field plane of the crystal: a setup with high overlap in the image plane may still suffer from low overlap in the far field or vice versa and this would translate into a decreased modal sensitivity. This overlap requirement has a central role in the derivation of eq. \eqref{Lgen}, which is based on the argument that the angular spread of a generated mode cannot exceed the natural spread of the down-conversion cone. In the next sections we will define an image plane bandwidth and a far field bandwidth and, as we shall show, there is a natural way of combining the two. This geometrical result is strongly supported by the more complicated analytic result, which we evaluate numerically for a comparison in figure 3.

\section{Analytical treatment}
A direct calculation of the measurement bandwidth needs to consider the overlap between the SPDC state and a pair of joint detection modes \cite{Torres:2003,Miatto:2011}. This yields a series of complex measurement amplitudes $\{C_\ell\}$ where $\ell$ labels each value of the OAM that was measured. The measured Schmidt number (or the measurement bandwidth) is therefore given by the measure $K$ applied to the set of projection probabilities
\begin{align}
K(\{P_\ell\}),\quad\mathrm{where}\quad P_\ell=|C_\ell|^2.
\label{MB}
\end{align}
We seek to evaluate this quantity for a Gaussian pump laser, taking full account of the sinc phase-matching term.  In this way we extend the regime of validity of earlier calculations. 

We consider the measurement modes for the signal and idler fields to be a pair of Laguerre-Gaussian modes.  
The LG modes are characterised by two integers $\ell$ and $p$ and a real positive number $w$, which represent the OAM quantum number, the radial quantum number and the Gaussian modal width, respectively.  For simplicity, we set $p=0$, which limits
our analysis to modes with a single bright ring in the transverse plane.  Many of our experiments are designed to detect $p=0$ modes with a higher efficiency, moreover, than higher-order modes.  We note however, that modes with non-zero 
$p$ are produced in the SPDC process \cite{Miatto:2011} and, indeed, it is these that makes it possible
to observe entanglement of three-dimensional vortex knots in SPDC \cite{RomeroPRL11}.   

The SPDC wave function $\psi(\mathbf{q}_i,\mathbf{q}_s)$, in momentum space, is written in the following way, where the subscripts $s$ and $i$ refer to signal and idler modes \cite{Torres:2003}:
\begin{align}
 \psi(\mathbf{q}_i,\mathbf{q}_s)=Ne^{-\frac{w_p^2}{4}|\mathbf{q}_i+\mathbf{q}_s|^2}\mathrm{sinc}\left(\frac{L}{4k_p}|\mathbf{q}_i-\mathbf{q}_s|^2\right) \, .
\end{align}
Here $\mathbf{q}$ is the transverse component of the momentum vector $\mathbf{k}$, $w_p$ is the pump width, $L$ is the crystal thickness, $k_p$ is the wave vector of the pump. The first term corresponds to the transverse wavevector 
components of the pump, while the second term represents the phase-matching imposed on the down-conversion process by
the nonlinear crystal.

We consider each detection mode to be an LG mode with radial quantum number $p=0$. In polar coordinates ($\rho,\varphi$) in momentum space it has the form
\begin{align}
 LG_\ell(\rho,\varphi)=\sqrt{\frac{w^2}{2\pi|\ell|!}}\left(\frac{\rho w}{\sqrt2}\right)^{|\ell|}e^{-\frac{\rho^2w^2}{4}}e^{i\ell\varphi}.
\end{align}
The projection amplitude is therefore calculated by evaluating the overlap integral of $\psi$ with two LG modes of opposite OAM (because of angular momentum conservation) \cite{MairNat01,Sonja:2002,Torres:2003}. The result is found to be
\begin{align}
 C_\ell^{L_R,\gamma}=\frac{\mathcal{N}}{L_R} \left(\frac{2\gamma^2}{1+2\gamma^2}\right)^{|\ell|}\left[\xi^{|\ell|+1}\Phi_\ell^{L_R,\gamma}-\Phi_{\ell}^{0,\gamma}\right] \, .
 \label{analytic}
\end{align}
We note that the first term in brackets corresponds to that obtained previously \cite{Miatto:2011}, specialised to equal signal and idler widths and $p=0$ modes.  Here the function $\Phi_{\ell}^{L_R,\gamma}$ is the Lerch transcendent function of order $(1,|\ell|+1)$ and argument $-2\gamma^2\xi$ \cite{watson1995treatise}:
\begin{align}
\Phi_{\ell}^{L_R,\gamma}=\Phi(-2\gamma^2\xi,1,|\ell|+1),\qquad \xi=\frac{i+L_R}{i-2\gamma^2L_R}.
\end{align}
Note that $\xi=1$ for $L_R=0$.

Once $L_R$ and $\gamma$ are specified, the amplitudes $C^{L_R,\gamma}_\ell$ are to be used in eq.~\eqref{MB}, in order to calculate the measurement bandwidth. The dependence of the projection amplitudes on a transcendent function makes further analytical calculation difficult, and a numerical approach has to be employed. However, as the tails of the distribution of projection probabilities have a slow decay and therefore an effect on the width even at high $|\ell|$, the numerical approach is slow, if an accurate result is sought.

\begin{figure}[ht*]
\begin{center}
\includegraphics[width=0.49\textwidth]{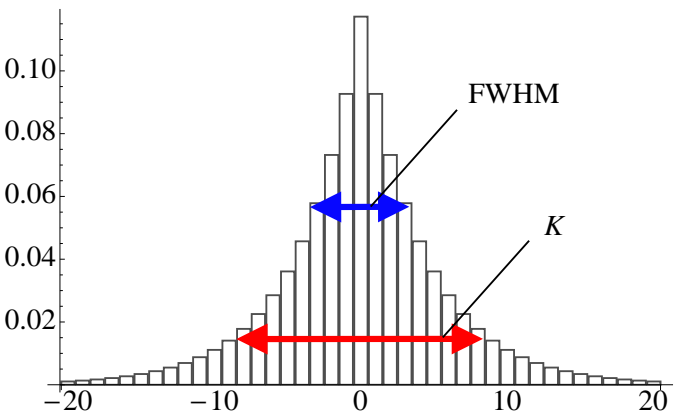}
\caption{\label{example}An example of a distribution of $|C_\ell^{L_R,\gamma}|^2$ for $L_R=0.001$ and $\gamma=2$, obtained by calculating numerically the projection amplitudes between $\ell=-20$ and $\ell=20$. The FWHM and the measurement bandwidth $K$ are shown in blue and red, respectively. Note that K exceeds the FWHM by about 2.5 times, giving an effective mode number 
of about 17 in this case.}
\label{fig:figure1}
\end{center}
\end{figure}

In figure \ref{fig:figure1} we give the probabilities for the angular momentum values $\ell$ for $L_R=0.001$ and $\gamma=2$. In this parameter range
existing analytical expressions provide an excellent approximation \cite{Torres:2003,Miatto:2011}.

\section{Geometrical argument}
In this section we find an upper (and therefore lower) bound for the generated OAM values, and for the measured OAM values. The measurement bandwidth that we calculate from such bounds matches the analytic result of the previous section and therefore allows to avoid calculating numerically the distribution of projection probabilities. 

\begin{figure}[ht]
\begin{center}
\includegraphics[width=0.48\textwidth]{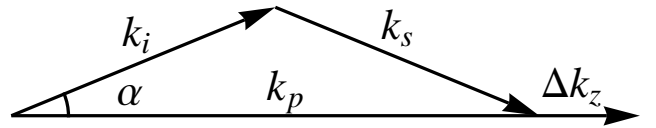}
\end{center}
\caption{The relation between $\alpha$ and $\Delta k_z$ sets a natural upper bound to $\alpha$ for near-collinear emission.}
\label{triangle}
\end{figure}

The phase-matching efficiency of the down-conversion process depends upon the axial mismatch $\Delta k_z$ in wave vectors of the pump, signal and idler fields, and it is given by $\mathrm{sinc}^2\left(L\Delta k_z/2\right)$.  When optimised for degenerate, near-collinear phase-matching, the signal and idler output is obtained over a narrow range of angles, $\alpha$, for which $L\Delta k_z \lesssim\pi$. With reference to figure \ref{triangle}, for small $\alpha$ (which corresponds to being near to collinearity) we can write
\begin{align}
\Delta k_z\simeq\frac{\alpha^2k_p}{2}.
\end{align}
It follows, therefore, that the allowed values of $\alpha$ are bounded from above:
\begin{align}
\alpha^2\lesssim\frac{2\pi}{k_pL}.
\end{align}

For Laguerre-Gaussian modes, in the paraxial regime, we can define an effective local wavevector associated with the gradient of the phase.  The helical form of the wavefronts gives rise to an angular spreading of these such that at a distance $r$ from the mode axis, the angular spread is $\beta\simeq\ell/k r$ \cite{PadgettOC95}, which can be interpreted as the local spreading angle from the optical axis.
The natural restriction on $\alpha$ imposed by the phase matching therefore sets a limit $\beta\lesssim\alpha$ on the efficiency of production of the OAM carrying beams, imposing a restriction on the generated OAM bandwidth. Such restriction is a natural consequence of the fact that a generated mode cannot be more divergent than the down-conversion cone. The relation $\beta\lesssim\alpha$, using the definitions and bounds given above for $\beta$ and $\alpha$, can be rewritten as
\begin{align}
\ell\lesssim r\sqrt{\frac{\pi k_p}{2L}} \, ,
\label{Lgen}
\end{align}
where we have made the approximation that $k_{s,i} \approx k_p/2$.
This relation is the starting point to calculate the generation bandwidth and for the analysis in the far field of the image plane of the crystal.

\subsection{Generation bandwidth}
The beam size can be no bigger than that of the pump beam, i.e. $r\lesssim w_p$. Applying this bound to eq.~\eqref{Lgen} we obtain an upper bound for the generated OAM value: 
\begin{align}
\ell_{\mathrm{gen}}\lesssim w_p\sqrt{\frac{\pi k_p}{2L}}=\sqrt{\frac{\pi }{L_R}}.
\end{align}
It follows, therefore, that the generation bandwidth is
\begin{align}
K_\mathrm{gen}=1+2\sqrt{\frac{\pi}{L_R}}.
\label{kgendef}
\end{align}
This number represents the effective number of entangled OAM modes generated by the source obtained by removing the $p=0$ restriction (as we are applying such restriction only to the measurement bandwidth). Equivalently, it can be thought of as the bandwidth obtained by removing the restriction on $\gamma$, i.e. if one does detect $p=0$, but with any $\gamma$. This way of thinking about $K_\mathrm{gen}$ can be helpful, as it relates to a measurement scheme. The relation between $K_\mathrm{gen}$ and the total Schmidt number $K$ or its azimuthal part $K_\mathrm{az}$ \cite{VanExter2006} is not straightforward, because $K_\mathrm{gen}$ can be thought of in terms of a measurement with any value of $\gamma$.

\subsection{Image plane bandwidth}

As anticipated in section 2, to calculate the measurement bandwidth we need to consider the overlap of the generated field with the detection modes in the image plane of the crystal and in its far field. Intuitively, a detection system which has a good overlap in the image plane, but that detects light that only comes from a narrow spread of directions would restrict the measured bandwidth. A similar restriction would occur for one that has a good overlap with the typical incoming angles of LG beams, but that has a poor overlap with the intensity in the image plane. It is clear that in order to optimise a detection system, both these quantities have to be taken into account.

To calculate the overlap in the image plane it suffices to note that a $p=0$ Laguerre-Gaussian mode with OAM number $\ell$ and width $w$ has its maximum intensity at a radius 
\begin{align}
r=w\sqrt{\frac{\ell}{2}}.
\end{align}
For efficient conversion of pump to signal and idler we require that the pump, single and idler beams should all overlap, giving a restriction on the maximum size of the down-converted beams ($r_{s,i}\lesssim w_p$) and hence an upper bound to the value of OAM in the plane of the crystal corresponding to 
\begin{align}
r_{s,i}=w_{s,i}\sqrt{\frac{\ell}{2}}\lesssim w_p.
\label{NFrestriction}
\end{align}

In terms of $\gamma$, this gives an upper bound of the value of the OAM in the plane of the crystal:
\begin{align}
\ell_\mathrm{ip}\lesssim2\gamma^2
\label{detband}
\end{align}
and hence an image plane bandwidth
\begin{align}
K_\mathrm{ip}=1+4\gamma^2 \, .
\end{align}

\subsection{Far field bandwidth}
It is clear that in the far field of the plane of the crystal, instead of a real space argument, we need to use the angular relationship $\beta\lesssim\alpha$, expressed in \eqref{Lgen}, where we apply the restriction for the maximum width of the detection modes given in \eqref{NFrestriction}:
\begin{align}
\ell\lesssim w_{s,i}\sqrt{\frac{\ell}{2}} \sqrt{\frac{\pi k_p}{2L}}.
\end{align}
From which, replacing $w_{s,i}$ with $w_p/\gamma$, we obtain an upper bound of the value of the OAM in the far field of the plane of the crystal:
\begin{align}
\ell_\mathrm{FF}\lesssim\frac{\pi}{2\gamma^2L_R}
\label{resgen}
\end{align}
and therefore a far field bandwidth
\begin{align}
K_\mathrm{FF}=1+\frac{\pi}{\gamma^2L_R}.
\end{align}

\subsection{Measurement bandwidth}
If $K_{\mathrm{ip}}$ and $K_{\mathrm{FF}}$ are very different from each other, the resulting measurement bandwidth is given by the smaller of the two. For cases where the bandwidths are similar it is sensible to combine them. The convolution of two normal distributions of widths $k$ and $k'$ gives a normal distribution of width $(k^{-2}+k'^{-2})^{-1/2}$. Similarly, we can get an estimate of the total measurement bandwidth by considering the convolution of two normal distributions of widths $K_\mathrm{ip}$ and $K_\mathrm{FF}$. The bandwidth of the resulting distribution is
\begin{align}
K&= \left(K_\mathrm{ip}^{-2}+K_\mathrm{FF}^{-2}\right)^{-1/2}\nonumber\\
&=\left(\left(1+4\gamma^2\right)^{-2}+\left(1+\frac{\pi}{\gamma^2L_R}\right)^{-2}\right)^{-1/2}.
\label{milesmath}
\end{align}

\section{Analysis of the results}

\begin{figure}[ht*]
\begin{center}
\includegraphics[width=0.49\textwidth]{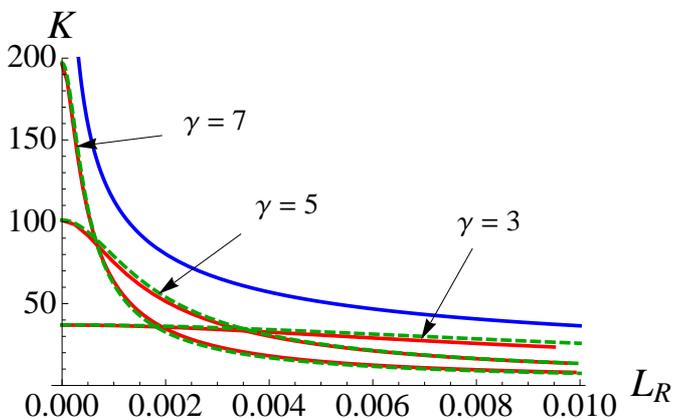}
\end{center}
\caption{\label{thegraph}(color online) The blue line (uppermost) is the generation bandwidth defined in \eqref{kgendef}, the green curves (dashed) are calculated from our analytical treatment , and the red curves (solid) are the result of our geometrical argument. 
}
\end{figure}

For a comparison between the analytic and geometric arguments, we calculate the width of the distribution given by the modulus squared of the coefficients in \eqref{analytic} and compare it to \eqref{milesmath}. In figure \ref{thegraph} we plot the two bandwidths as functions of $L_R$ for $\gamma=3$, $\gamma=5$ and $\gamma=7$. The solid curves (red online) represent the measurement bandwidth calculated from the numerical evaluation of he analytical model. The dashed curves (green online) are the same bandwidths calculated with our geometrical argument. The uppermost solid line (blue online) is the generation bandwidth. Note that to achieve high dimensional entanglement the crystal length should be a small fraction of the Rayleigh range.

We see that the geometrical argument is in excellent agreement with the numerical evaluation of our analytical result. The effect of increasing $\gamma$ yields a higher measurement bandwidth for very small values of $L_R$, but for large enough values of $\gamma$ and for fixed $L_R$, the measurement bandwidth eventually drops. Therefore it reaches a maximum value for a particular crystal length. Under all conditions the measurement bandwidth never reaches that of the generation bandwidth, because we are restricting the measurement to modes with $p=0$. Note, however, that the full generation bandwidth does
not arise explicitly from additional values of the OAM but rather from entanglement in the radial quantum number $p$.

Differentiation of eq.~\eqref{milesmath} with respect to the crystal length gives an estimate of the value of $\gamma$ corresponding to the highest measurement bandwidth for a given $L_R$.  In this way we find
\begin{align}
\gamma_{\mathrm{opt}}\approx\sqrt[4] {\frac{\pi}{4L_R}}.
\label{gammaopt}
\end{align}
It is worth noting that for such value of $\gamma$ we have that $K_\mathrm{ip}=K_\mathrm{FF}=K_\mathrm{gen}$, where $K_\mathrm{gen}$ is defined in \eqref{kgendef}. Therefore in the optimal case we have $K=K_\mathrm{gen}/\sqrt{2}$.

We define short crystal lengths as $L_R\ll\pi/4\gamma^4$, for which the generation bandwidth is large, meaning that the measurement bandwidth is dominated by the image plane overlap of the detection modes with the pump.  This gives a measurement bandwidth of
\begin{align}
K\approx K_{\mathrm{ip}}=1+4\gamma^2.
\end{align}
Note that this short crystal limit is characterised by an independence of $K$ on the crystal length. In fact, it can be seen in figure \ref{thegraph} that the leftmost part of the measurement bandwidth curves is flat (for the $\gamma=7$ curve this is not visible in this plot, but the slope of eq.~\eqref{milesmath} near the origin is zero for any $\gamma$), and that the range of values of $L_R$ over which they stay flat is inversely proportional to $\gamma^4$.
For much longer crystals, $L_R\gg\pi/4\gamma^4$, the measurement bandwidth, as modified by the limiting overlap in the far field, becomes dominant, giving
\begin{align}
K\approx K_{\mathrm{FF}}=1+\frac{\pi}{L_R \gamma^2}.
\end{align}

\begin{figure}[ht*]
\begin{center}
\includegraphics[width=0.49\textwidth]{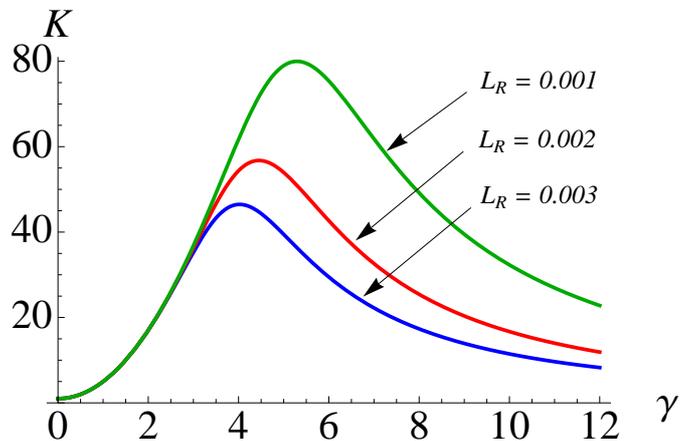}
\caption{\label{kgamma}An example of a measurement bandwidth as a function of $\gamma$ for three different values of $L_R$: 0.001, 0.002 and 0.003 (highest to lowest).}
\end{center}
\end{figure}

In figure \ref{kgamma} we plot three different curves, that describe the value of the measurement bandwidth as a function of $\gamma$, for three different values of $L_R$. Note that for each choice of $L_R$ there is always an optimal value of $\gamma$ which maximises $K$, and it corresponds to the optimal value given in \eqref{gammaopt}.

It is not an easy matter to determine the requisite parameters for existing experiments.  Most our own experiments, however, 
correspond to values of $\gamma$ in the range $1.5$ up to about $4$.  In order to achieve higher degrees of entanglement in
OAM, corresponding to larger Schmidt number, our analyses suggest that it would be desirable to press towards higher values of $\gamma$.

\section{Conclusions}

We have shown two parameters determine the OAM bandwidth for entangled states produced by parametric down-conversion. These parameters are the ratio of the widths of pump and detection modes $\gamma=w_p/w_{s,i}$, and the crystal thickness normalised to the Rayleigh range of the pump $L_R=L/z_r$. 

A simple geometrical argument approximates the analytical results extremely well and allows us to suggest what needs 
to be adjusted in order to enhance the dimensionality of the entanglement. We have restricted our analysis to a  detection system that is sensitive to the LG $p=0$ modes only.  It is for this reason that the measurement bandwidth can never reach that of the generation bandwidth for any combination of parameters.  It is possible, however, to identify an optimum value of $\gamma$ to maximise the measurement bandwidth for any normalised crystal length $L_R$. 

\section*{Acknowledgements}This work was supported by the UK EPSRC.
We acknowledge the financial support of the Future and Emerging
Technologies (FET) program within the Seventh Framework Programme
for Research of the European Commission, under the FET Open grant
agreement HIDEAS number FP7-ICT-221906.
This research was supported by the DARPA InPho program through the US Army Research Office award W911NF-10-1-0395.
SMB and MJP thank the Royal Society and the Wolfson Foundation for financial support.  SMB thanks Alison Yao for 
her invaluable assistance with this manuscript.

\bibliographystyle{epj}
\bibliography{Bandwidths}

\end{document}